\documentclass[11pt,a4paper]{article}
\usepackage{amsmath, amsbsy, amsthm, enumerate, latexsym, subfigure}
\usepackage{graphicx}
\usepackage{amsmath}
\newcommand{\ra}{\rightarrow}

\newcommand{\lra}{\longrightarrow}

\newcommand{\ga}{\gamma}                         %\ga=\gamma
 \numberwithin{equation}{section}

\newcommand{\bs}{\bigskip}

\def\expec{\text{\rm I\kern-2pt E}}
\def\ex{\expec}
\def\prob{\text{\rm I\kern-2pt P}}
\def\pr{\prob}

\begin{document}

%\baselineskip=16pt

%\hfill \today \bs

\title{\Large Systematic assessment of the expected length, variance and distribution of Longest Common Subsequences} \bs

\author{
Kang Ning\\Qingdao Institute of BioEnergy and Bioprocess Technology\\Chinese Academy of Sciences\\Qingdao, Shandong, China, 266101\\ningkang@qibebt.ac.cn
\and
Kwok Pui Choi\\Department of Statistics and Applied Probability\\National University of Singapore\\Singapore, 48105\\stackp@nus.edu.sg
}

\date{\today}

\maketitle

\begin{abstract}  % enter abstract text below this line

The Longest Common Subsequence (LCS) problem is a very important problem in mathematics, which has a broad application in scheduling problems, physics and bioinformatics. It is known that the given two random sequences of infinite lengths, the expected length of LCS will be a constant. however, the value of this constant is not yet known. Moreover, the variance distribution of LCS length is also not fully understood. The problem becomes more difficult when there are (a) multiple sequences, (b) sequences with non-even distribution of alphabets and (c) large alphabets. This work focus on these more complicated issues. We have systematically analyze the expected length, variance and distribution of LCS based on extensive Monte Carlo simulation. The results on expected length are consistent with currently proved theoretical results, and the analysis on variance and distribution provide further insights into the problem.

\end{abstract}  % abstract text must end above this line

\section{\bf Introduction}
\label{sec:1}

\subsection{The problem}
\label{sec:problem}

Given a string of characters $A = a_1a_2...a_n$, a subsequence of $A$ is a string which can be obtained by deleting some of the characters in $A$. Given another string $B = b_1b_2...b_n$, the longest common subsequence (LCS) of A and B (LCS(A,B)) is the longest sequence which is a subsequence of both A and B.

Let $L_n$ be the length of the longest common subsequence of two random binary sequences (A and B) of length $n$. It is known by subadditivity that the there exists a $\gamma$, so that expected value of $L_n$, $\lim_{n\to\infty}(E[L_n]/n) = {\gamma}$ \cite{chvtal_longest_1975}, this value is called Chvatal-Sankoff constant. It is tempting to compute the value of $\gamma$ when the choices are made with equal probability from an alphabet set of size $q$, and there is special charm to $q=2$, the case of the longest common subsequence of two binary sequences.

\subsection{The objectives}
\label{sec:objective}

There is extensive research on the determination of $\gamma_2$. However, the exact value of $\gamma_2$ is not known. But there are researches on computation of the lower and upper bounds of the value of $\gamma_2$. Moreover, research on the expected length of LCS of more that two sequences ($\gamma_i$ with $i>2$) is rare.

Another interesting problem is the analysis of the variances of the longest common subsequence. The problem concerns the growth rate of the variance, and the basic issue is to decide if the variance grows linearly with $n$. This problem has not been studied extensively by researchers, and there are only some intriguing suggestions.

In this paper, we have focused on this ratio, $\gamma$. We will perform systematic assessment of the length of LCS under differnt settings.

\begin{enumerate}
\item Experimental design and datasets selection. This include different settings with (a) multiple sequences, (b) sequences with non-even distribution of alphabets and (c) large alphabets.
\item Extensive analysis on expected length, variance and distribution of LCS length under these different settings.
\end{enumerate}

Table \ref{Current_Analysis} shows the areas in which we will analyze, as well as current results in these areas. There are already some simulation work on the length and variants of $L_n$ for LCS \cite{bundschuh_high_2001,boutet_de_monvel_extensive_1999}. However, current work mainly focused on two completely random sequence, and the systematic analysis on multiple sequences based on larger alphabet and non-even distributed alphabet is rare. Moreover, we will  assessment of LCS algortihms based on these systematic analysis results as benchmark.

\begin{table}[htbp]
\caption{\label{Current_Analysis} Current analysis results on LCS with different settings. Our analysis is mainly based on Monte-Carlo simulation.}
\begin{center}
\begin{tabular}{|l|l|p{3cm}|p{3cm}|p{3cm}|}
\hline
 & & $\ga$ & Variance & Distribution \\
\hline
No. of seqs & 2 & theoretical lower and upper bounds & simulation analysis with few seqs & simulation analysis with few seqs \\
\hline
 & multiple &  &  &  \\
\hline
Alphabet size & binary & theoretical lower and upper bounds & simulation analysis with few seqs & simulation analysis with few seqs \\
\hline
 & $\Sigma>2$ & simulation analysis with few seqs & simulation analysis with few seqs &  \\
\hline
Distribution of alphabets & random & theoretical lower and upper bounds & simulation analysis with few seqs &  \\
\hline
 & non-random &  &  &  \\
\hline
\end{tabular}
\end{center}
\end{table}

The analysis based on Monte Carlo method would be very complicated for some settings. For example, the analysis of LCS of multiple sequences based on large alphabet set is so complicated that analytical results based on simulations can only provide some bounds. 

In the following part, we will use the annotations as listed below:

\begin{description}

\item[$n$]: The length of the sequence.

\item[$\gamma_i^q$]: The ratio of expected length of LCS of $i$ sequences based on q alphabets, over sequence length. If the the alphabet size is  $q=2$, then the simple form $\gamma_i$ is used.

\item[$Var(|LCS|_i^q)$]: The variants of the length of LCS of $i$ sequences based on q alphabets.

\end{description}

\section{Literature Review}
\label{sec:2}

\begin{itemize}

\item Previous conjectures on $\gamma_2$ and $VAR(L_n)$

Previously there is a conjecture that $\gamma_2 = 2/(1+\sqrt{2}) \approx 0.828427$ \cite{chvtal_longest_1975, steele_probability_1997}, but it is not true \cite{lueker_improved_2003}. There is another conjecture \cite{steele_probability_1997} suggesting that $\gamma_2$ is closer to 0.81 than to $2/1+\sqrt{2}$. There is also conjecture \cite{alexander_rate_1994} based on simulations that $0.8079 \leq \gamma_2 \leq 0.8376$ with at least $95\%$ confidence.

As for the variance of $|LCS|$, $VAR(|LCS|)$. There is simulation evidence in \cite{waterman_introduction_1995} for the conjecture that for two sequences, it grows linearly with the length of the sequence $n$, and there is another conjecture \cite{chvtal_longest_1975} based on simulation that the variance was of much lower order. Recently, there is theoretical proof that for a constant $c>0$, $VAR(|LCS|)>cn$ if $\epsilon$ is taken small enough \cite{blasiak_longest_2004}. This reject the Chvatal-Sankoff conjecture that$VAR(|LCS|)=o(n^{2/3})$.

\item Computation of lower bound and upper bound of $\gamma_2$

Previously, there is a bound of $\gamma_2$ \cite{dan&ccaron;k_upper_1995,paterson_longest_1994} that $$0.773911 \le \ga _2 \le 0.837623$$.

The computation of the lower bound and upper bounds of $|LCS|$ is essentially the analysis of a Markov chain corresponding to a finite automaton that read pairs of strings.

For the computation of lower bound of $|LCS|$, the finite-state automata is built such that it can read input characters, left-to-right, from the string A and B, producing one character of output whenever it matches a  character from A and B. The behaviors of this automata can be modeled as a Markov chain, so that the expected number of matches output per character read can be computed. there are 931 states in the automata constructed in \cite{dan&ccaron;k_upper_1995,paterson_longest_1994}, and this is used to produce the lower bound of $\gamma \geq 0.773911$.

For the computation of upper bound of $|LCS|$, in \cite{dan&ccaron;k_upper_1995,paterson_longest_1994}, the authors pointed out that all upper bounds as of his writing use the following basic idea: Let $G(k, n)$ be the number of pairs of strings of total length $n$ whose LCS has length exactly $k$. Then a simple argument shows that if for some fixed $y$ and large $n$ we have $\sum^{n}_{i={\lceil yn \rceil}}G(i,n)=o(4^n)$, then we can conclude that $\gamma \leq y$. For a given string $S$ of length $k$, let an $n-collation$ of $S$ be any way of inserting additional characters into two copies of $S$ to form two strings $A$ and $B$ of total length $n$. Let $C_n(S)$ be the number of distinct $n-collations$ possible for string $S$. Then certainly $G(k,n) \leq \sum_{S\in{0,1}_k}C_n(S)$. The problem is that this bound may be weak: a given pair of strings may be counted many times, since many different strings may be an LCS. Thus a method that has been used to improve bounds is to try to restrict the eligible collations in such a way that this overcounting is minimized. To obtain the upper bounds, \cite{dan&ccaron;k_upper_1995,paterson_longest_1994} define a notion of a dominated collation, and use a counting argument based on an automaton that reads in match pairs and rejects when it can determine that the collation is dominated. A carefully constructed 52-state automaton yields a bound of $\gamma \leq 0.837623$.

Recently, there is a tighter bound \cite{lueker_improved_2003} that

$$0.788071 \le \ga _2 \le 0.826280$$.

The principle methods used in \cite{lueker_improved_2003}is similar to that used in \cite{dan&ccaron;k_upper_1995,paterson_longest_1994}. The main difference is that in \cite{dan&ccaron;k_upper_1995,paterson_longest_1994}, the researchers used carefully constructed automata, while in \cite{lueker_improved_2003}, the authors used automata with many states based on dynamic programming process.

For the computation of the lower bound of $\gamma_2$, the authors of \cite{lueker_improved_2003} generated the automata based on dynamic programming. Let $s$ and $t$ be arbitrary bit sequences of length $n$, let $v_m(s,t)$ be the maximum, for $0 \leq i \leq m$, of the expected length of the LCS of a pair of strings obtained by appending $i$ random characters to $s$ and $m - i$ random characters to $t$. Using the result of \cite{dan&ccaron;k_upper_1995,paterson_longest_1994} one easily sees that regardless of $s$ and $t$, $\gamma = lim_{n\rightarrow\infty}n^{-1}v_{2n}(s,t)$. A recurrence for the set of variables $v_n(s, t)$ can easily be obtained by dynamic programming formulas. The vector recurrence gives the lower bound.

For the computation of upper bounds, the researchers of \cite{lueker_improved_2003} used a method similar to \cite{dan&ccaron;k_upper_1995,paterson_longest_1994}, but they use dynamic programming techniques to control the overcounting.

Recently, there are two additional approaches for computing the bounds for $\gamma_2$. One is the Finite Width Model Sequence (FWM) for sequence comparison \cite{chia_finite_2004}. Another one is Bernoulli matching model based on analyses of the r-reach simplification \cite{blasiak_longest_2004}. The later one is interesting in that it has some inplicite linkage with heuristic algortihms for LCS problem.

\item Simulation results

There are already simulation works on length, variant and distribution of $|LCS|$ \cite{bundschuh_high_2001,boutet_de_monvel_extensive_1999}. In \cite{bundschuh_high_2001}, a well-defined asymptotic behavior of the difference $\gamma_2*n-E(|LCS|)$ , and a scaling difference in $VAR(|LCS|)-n^(2/3)$, are discovered. THe distribution of $(|LCS|-E(|LCS|)/\sqrt{Var(|LCS|)}$ is also empirically proven to be nearly normally distributed. In \cite{boutet_de_monvel_extensive_1999}, $\gamma_2*n\approx(a-b/W)$ is empirically proved to be the approximate value of $\gamma_2*n$, in which $W$ is the length of sequences, while $a$ and $b$ are constants.

However, current work mainly focused on two completely random binary sequence, and the systematic analysis on multiple sequences where alphabet are larger and non-random is rare.

\item Related theoretical results

There are quite a lot of theoretical results on th expected score of the multiple sequence alignment \cite{waterman_introduction_1995}. Since the problem of computation of LCS is actually the multiple sequence alignment problem with special matching scores. Namely, if we set insertion and deletion score to be 0, match score to be 1 and mismatch score to be $-\infty$, then on the same set of sequences, the optimal multiple sequences alignment score is equal to $|LCS|$ for these sequences. Therefore, we believe that the theoretical results for multiple sequence alignment can be helpful for our analysis on the expected length and variance of $|LCS|$.

From these related studies, we expect the linear growth of $|LCS|$ with respect to the length of sequences, as well as the large deviations from the mean of $|LCS|$.

\end{itemize}

\section{Approaches for Assessment}
\label{sec:3}

Use simulation (simulations up to $n=5000$.)

\section{Experiment settings and Results}
\label{sec:4}

The expected values, variances and distribution of $\gamma$ (especially $\gamma_2$) is of great interest to researchers (\cite{steele_probability_1997}). The computational results, either exact or simulated, are mentioned in some papers (\cite{chvtal_longest_1975,steele_probability_1997}) without much detail and analysis. In this part, we will describe our experimental settings and analysis results in details.

\subsection{Experiment setting and datasets}
\label{sec:settings}

Experiment setting and datasets can serve as benchmark for LCS analysis as well as for comparison of LCS algorithms.

We have made direct Monte-Carlo evaluation of $\gamma_2$ for (a) sequences with length $n$ from 1 up to cetain values, and (b) large set of random sequences.

\begin{enumerate}
\item For sequence lengths $1 \leq n \leq 20$, the instances for computation are all of the possible sequences for each $n$.
\item For sequence lengths $20 \leq n \leq 2000$, the number of instances for computation is $max(2^n, 10^5)$ for each $n$.
\item For sequence lengths $2000 \leq n \leq 10^4$, the number of instances for computation is $10^4$ for each $n$.
\item For sequence lengths $10^4 \leq n \leq 10^5$, the number of instances for computation is $10^2$ for each $n$. Since there are so few instances in this group, $\chi^2$ analysis are performed to analyze the extrapolated estimates, and the simulation errors are also estimated.
\end{enumerate}

For analysis of variances and distributions, we have also used these datasets. In category (1) datasets, since we have all the sequence for a specific length, the average, variants and distribution can be calculated exactly. In datasets from category (2)-(4), $\chi^2$ can be applied for estimation of these values. We sould mainly analyze these values based on Monte Carlo simulation.

\subsection{Generation of sequences datasets}
\label{sec:seq_generation}

to generate the sequences datasets, we have used the in-house random sequences generator.

(description)

\subsection{Analysis of sequences datasets}
\label{sec:seq_analysis}

Since this study will focus on systematic assessment of LCS problem, the datasets should be representative of the problem to be studied. Therefore, we have analyzed the properties of the sequences datasets.

The first and most important analysis is on the converge of randomly generated sequences over all different sequences of the specific length. This analysis is necessary since our analysis will be performed on dataset not only with short sequences, but also on datasets with very long sequences. For datasets with short sequences, all of the possible sequences can be enumerated. However, for datasets with long sequences, enumerating all sequences is not practical, and only a subset of these sequences can be used for analysis. A good subset should cover a significant portion ("coverage") of the full set of the possible sequences of the specified length. Therefore, analysis of the coverage of such subsets is very important.

We have analyzed the coverage of these sequences for dataset categories (2)-(4).

The next analysis is on the distribution of alphabet contents (definition) of among all of the sequences in each dataset. We have used random sequences generator to generate the sequences datasets in categories (2)-(4). However, the exact distribution of the alphabet contents should be analyzed after that to make sure that the sequences in the datasets are evenly distributed, or distributed according to the predefined distribution.

Then we have analyzed the distribution of alphabet contents (definition) at different positions in the sequences.

Results show that the datasets that we have generated for LCS problem analysis have high coverage, and have appropriate distributions.

\subsection{Short sequences}
\label{sec:short_seq}

Compute the actual values of $\ga_2$ for $n = 2 \sim 15$.

The results are shown in Table \ref{Exact}.

\begin{table}[htbp]
\caption{\label{Exact} Exact Length of LCS and Variances for n =2 to 15.}
\begin{center}
\begin{tabular}{|l|l|l|l|}
\hline
Length & $|LCS|$  & $\ga _2$ & Variance \\
\hline
2 & 1.125 & 0.5625 & 0.359375 \\
\hline
3 & 1.8125 & 0.604166667 & 0.46484375 \\
\hline
4 & 2.5234375 & 0.630859375 & 0.577575684 \\
\hline
5 & 3.24609375 & 0.64921875 & 0.685531616 \\
\hline
6 & 3.979980469 & 0.663330078 & 0.783290625 \\
\hline
7 & 4.721435547 & 0.674490792 & 0.876503408 \\
\hline
8 & 5.469116211 & 0.683639526 & 0.965354785 \\
\hline
9 & 6.221725464 & 0.691302829 & 1.050569264 \\
\hline
10 & 6.978439331 & 0.697843933 & 1.132237011 \\
\hline
11 & 7.738685608 & 0.703516873 & 1.210766569 \\
\hline
12 & 8.501921177 & 0.708493431 & 1.28666914 \\
\hline
13 & 9.267754078 & 0.71290416 & 1.360145849 \\
\hline
14 & 10.03585378 & 0.716846699 & 1.431442313 \\
\hline
15 & 10.80596581 & 0.720397721 & 1.50072875 \\
\hline
\end{tabular}
\end{center}
\end{table}

\begin{figure}[htbp]
\begin{center}
\includegraphics[height=6.18cm, width=10cm]{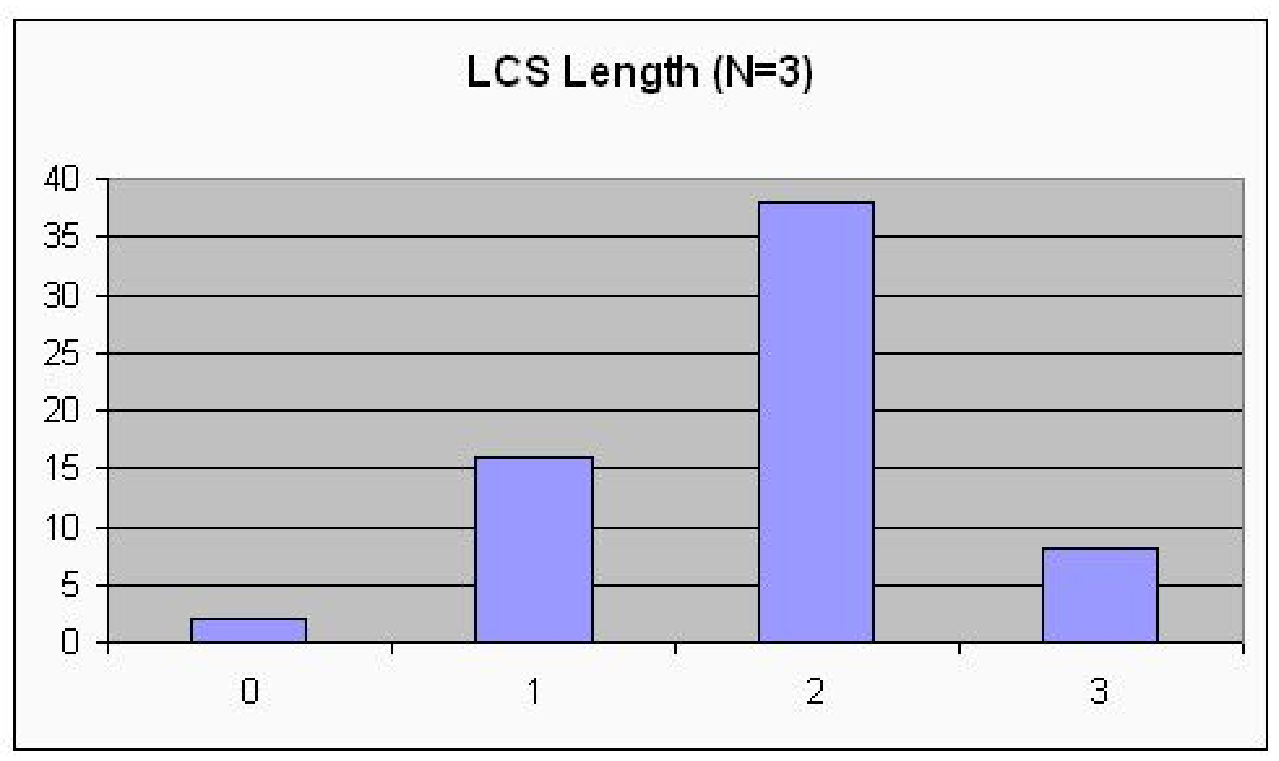}
\caption{\small \sl The Distribution of the lengths of LCS, n = 3.}
\label{fig:N3}
\end{center}
\end{figure}

\begin{figure}[htbp]
\begin{center}
\includegraphics[height=6.18cm, width=10cm]{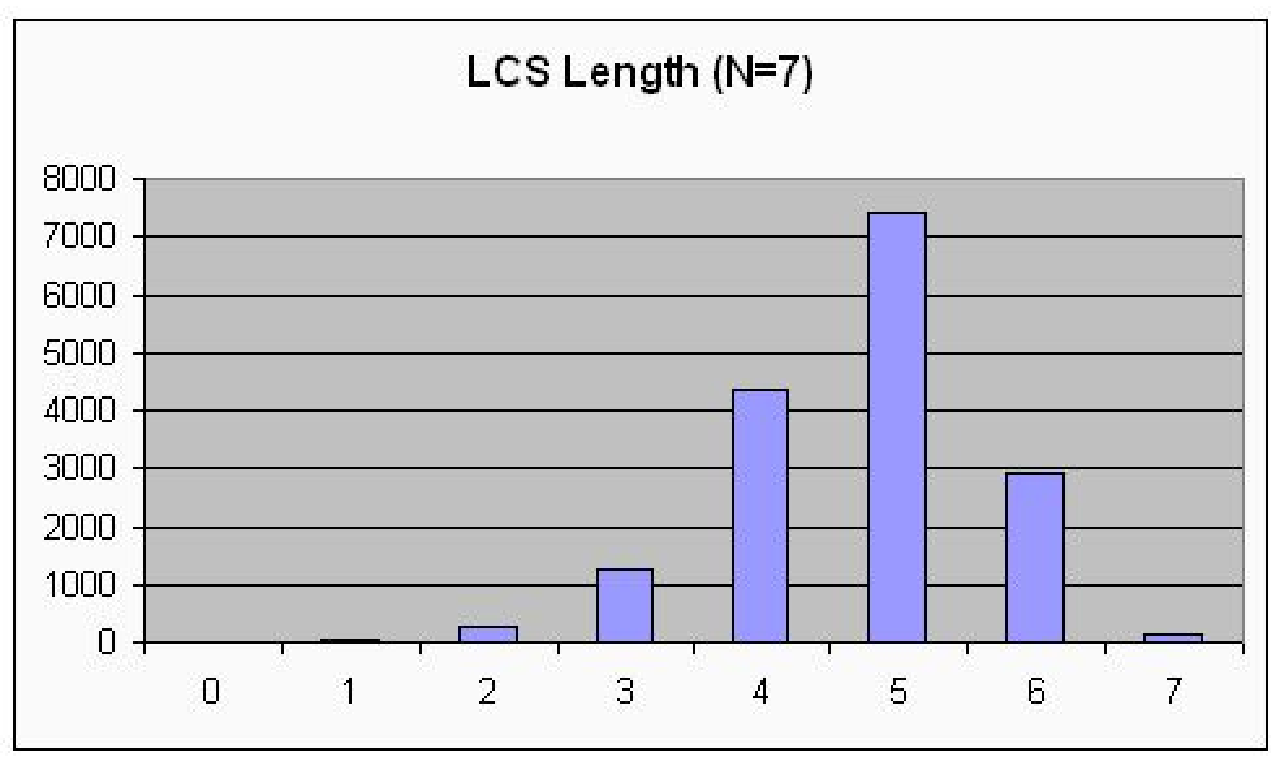}
\caption{\small \sl The Distribution of the lengths of LCS, n = 7.}
\label{fig:N7}
\end{center}
\end{figure}

\begin{figure}[htbp]
\begin{center}
\includegraphics[height=6.18cm, width=10cm]{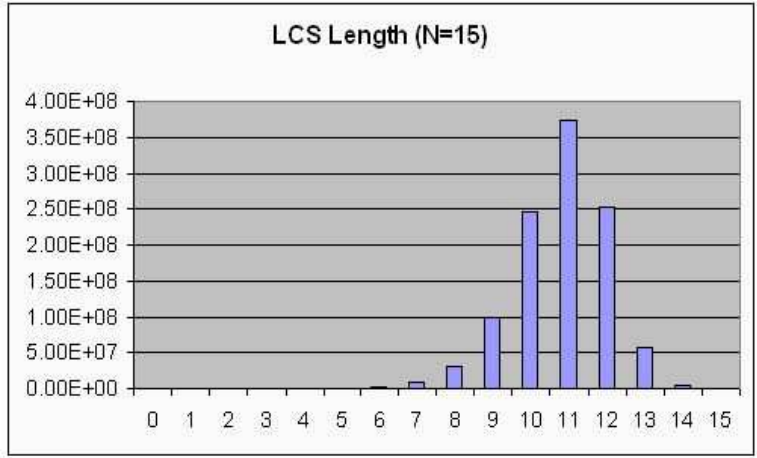}
\caption{\small \sl The Distribution of the lengths of LCS, n =15.} \label{fig:N15}
\end{center}
\end{figure}

The distributions of $|LCS|$ are interesting. Figures
(Figure \ref{fig:N3}, Figure \ref{fig:N7} and Figure
\ref{fig:N15}) illustrate the distribution of $|LCS|$
for $2 \leq n \leq 15$.

We observed that the lengths of the LCS are centered around $n-\delta$, and concentrated in the area $(n-\delta)\pm \delta$, where $\delta \simeq int(n/4)+1$ for $2 \leq n \leq 15$.

\subsection{Long sequences}
\label{sec:long_seq}

The exact expected values and variances of $|LCS|$ are not possible for datasets with long sequences. To assess the average value of $|LCS|$, its variance and the value of $\gamma_2$ for long sequences, the Monte-Carlo simulation method can be applied for the approximate values. For the simulations, a few (much less than the total number of sequences of specific length) random sequences are generated, and the average and variances of $|LCS|$ are computed based on these sequences.

The Monte Carlo simulation results for a few long sequences are described in \cite{chvtal_longest_1975}. Based on their results, the authors conjectured that $\gamma_2 > 4/5$ and that the variance of $|LCS|$ is $o(n^{2/3})$.

In our experiments, to assess the accuracy of simulations, we have first performed the simulation on short sequences, and compared with the exact results.

\begin{figure}[htbp]
\begin{center}
\includegraphics[height=6.18cm, width=10cm]{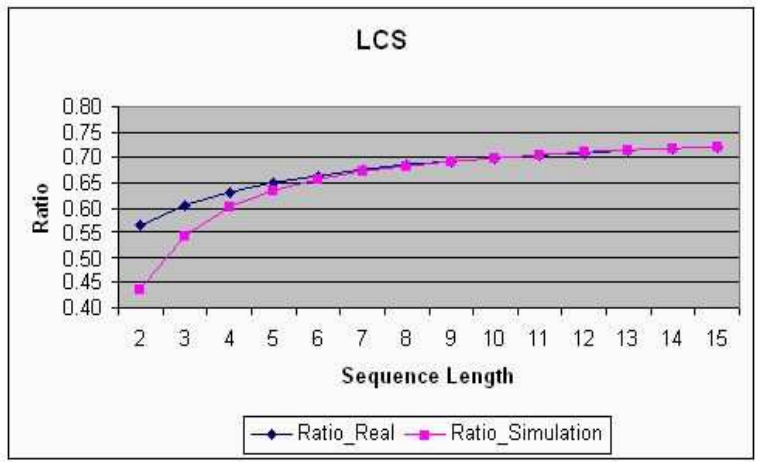}
\caption{\small \sl The comparison of the exact and simulated
results of $\gamma_2$.} \label{fig:Compare_Ratio}
\end{center}
\end{figure}

\begin{figure}[htbp]
\begin{center}
\includegraphics[height=6.18cm, width=10cm]{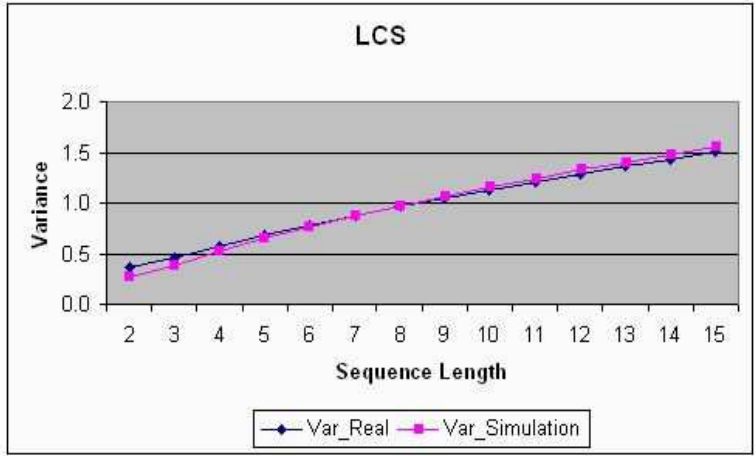}
\caption{\small \sl The comparison of the exact and simulated
results of $Var(|LCS|)$.} \label{fig:Compare_Var}
\end{center}
\end{figure}

For the simulation of specific sequence length, each dataset contains $2^8$ sequences, and we have repeated the simulation on $2^{10}$ different datasets to average out the values. The comparisons are illustrated in Figure \ref{fig:Compare_Ratio} and Figure \ref{fig:Compare_Var}.

From the comparisons, we have observed that the simulations are quite accurate. We can represent the difference of exact and simulated results of $\gamma_2$ and variances as

\begin{equation}
|\gamma_2 - \hat{\gamma_2}| \leq \varepsilon_{\gamma_2}
\end{equation}

\begin{equation}
|Var(|LCS|) - \hat{Var(|LCS|)}| \leq \varepsilon_{Var(|LCS|)}
\end{equation}

where $\gamma_2$ and $\hat{\gamma_2}$ are exact and simulated values of $\gamma_2$, and $\varepsilon_{\gamma_2}$ is the difference. For variances, $Var(|LCS|)$ and $\hat{Var(|LCS|)}$ are exact and simulated values of $Var(|LCS|)$, and $\varepsilon_{Var(|LCS|)}$ is their difference.

The observation from short sequences is that the values of $\varepsilon_{\gamma_2}$ is less than 0.01, and the values of $\varepsilon_{Var(|LCS|)}$ is less than 0.1. This prove very confident simulation results. Based on these comparisons, we are quite confident about the simulation results on long sequences.

For long sequences, we have computed the approximated values of $\ga _2$ for $n = 16\sim25$, 50, 100, 150, 200, 250, 300, 350, 400, 450, 500, 1000, 1500, 2000, 2500, 3000, 3500, 4000, 4500, 5000 based on simulated sequences datasets.

For simulated sequences dataset, we have set each dataset containing $2^8$ sequences, and we have set the number of such datasets according to the length of the sequences ($2^{10}$ for $n \le 25$, and $2^7$ for $n > 25$).

For experiments with very long sequences ($n \geq 1000$), each of the datasets contains $2^4$ sequences, and there are $2^4$ datasets.

The results are shown in Table \ref{Simulation}.

\begin{table}[htbp]
\caption{\label{Simulation} Approximate $\ga _2$ and
Variances for large n.}
\begin{center}
\begin{tabular}{|l|l|l|}
\hline
Length & $\ga _2$ & $Var(|LCS|)$ \\
\hline
16 & 0.725 & 1.633 \\
\hline
17 & 0.728 & 1.715 \\
\hline
18 & 0.730 & 1.780 \\
\hline
19 & 0.733 & 1.854 \\
\hline
20 & 0.735 & 1.923 \\
\hline
21 & 0.737 & 1.994 \\
\hline
22 & 0.739 & 2.058 \\
\hline
23 & 0.741 & 2.128 \\
\hline
24 & 0.743 & 2.194 \\
\hline
25 & 0.744 & 2.269 \\
\hline \hline
50 & 0.760 & 4.680 \\
\hline
100  & 0.783 & 7.028 \\
\hline
150  & 0.789 & 10.814 \\
\hline
200  & 0.793 & 15.059 \\
\hline
250  & 0.796 & 19.970 \\
\hline
300  & 0.797 & 25.482 \\
\hline
350  & 0.799 & 31.603 \\
\hline
400  & 0.800 & 38.621 \\
\hline
450  & 0.801 & 46.228 \\
\hline
500  & 0.802 & 54.306 \\
\hline \hline
1000 & 0.806 & 174.768 \\
\hline
1500 & 0.818 & 311.617 \\
\hline
2000 & 0.819 & 545.767 \\
\hline
2500 & 0.8201 & 845.373 \\
\hline
3000 & 0.8204 & 1213.201 \\
\hline
3500 & 0.8210 & 1638.397 \\
\hline
4000 & 0.8211 & 2136.979 \\
\hline
4500 & 0.8214 & 2694.796 \\
\hline
5000 & 0.8215 & 3323.126 \\
\hline
\end{tabular}
\end{center}
\end{table}

\begin{figure}[htbp]
\begin{center}
\includegraphics[height=6.18cm, width=10cm]{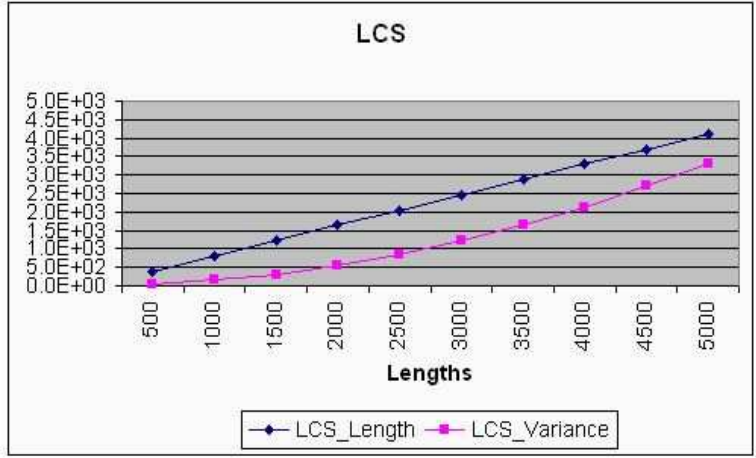}
\caption{\small \sl The lengths and $Var(|LCS|)$ for sequences
with length $\geq 500$.} \label{fig:N500}
\end{center}
\end{figure}

\begin{figure}[htbp]
\begin{center}
\includegraphics[height=6.18cm, width=10cm]{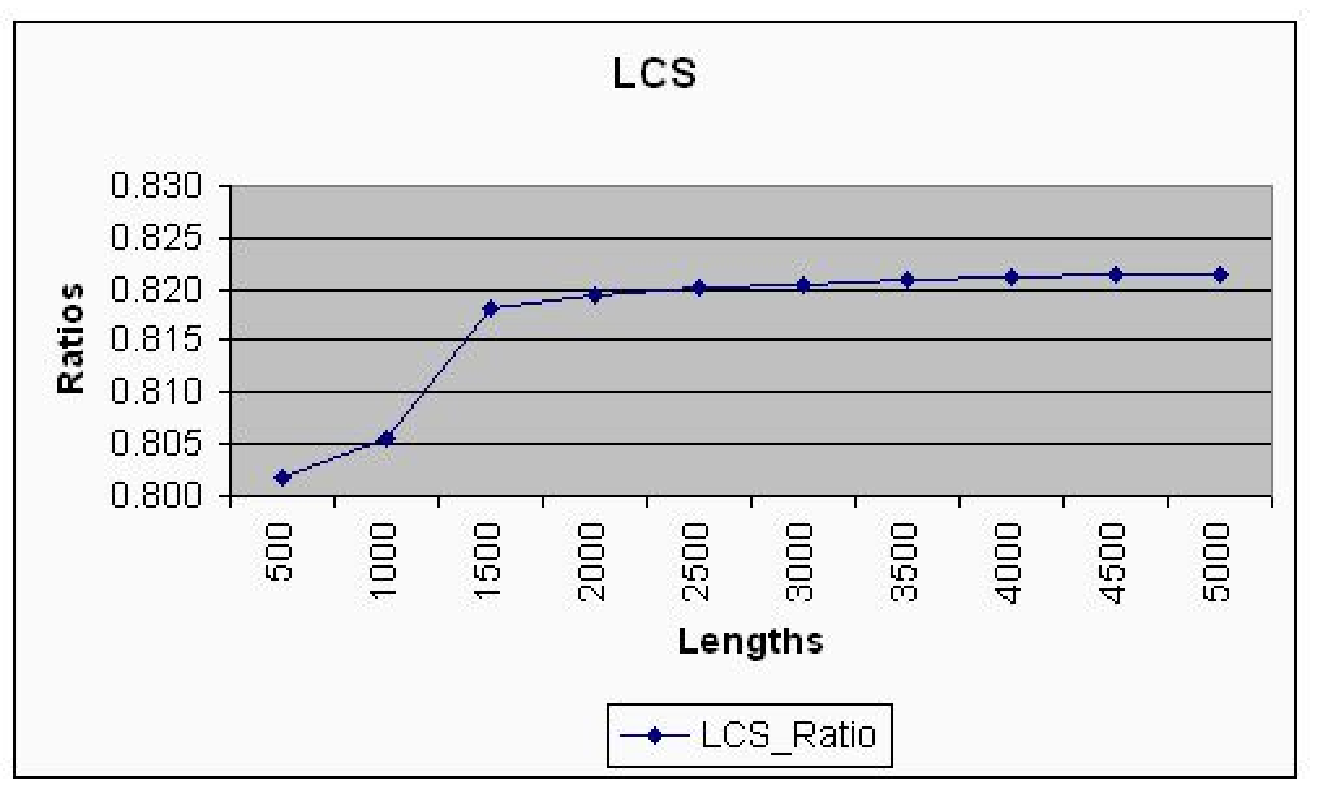}
\caption{\small \sl The approximate values of $\gamma_2$ for
sequences with $n \geq 500$.} \label{fig:N500_ratio}
\end{center}
\end{figure}

An illustration of the values of $|LCS|$ and $Var(|LCS|)$, and approximate values of $\gamma_2$ are illustrated in Figure \ref{fig:N500} and Figure \ref{fig:N500_ratio}.

For the simulated sequences datasets, the ratio $\ga _2$ reaches lower bound (0.788071) when the sequences length is around 150, and there is still some gap to the upper bound (0.826280) when the sequences length is up to 5000.

The variances of $|LCS|$ increase with the length of the sequences in a supper-linear manner. The rate at which the variances increase is higher than the rate at which the average $|LCS|$.

Based on our simulation results, we are confident to conjecture that $\gamma_2 > 0.82$. Combined with the upper bound of 0.826280 given by Lueker et.al. \cite{lueker_improved_2003}, the actual value of $\gamma_2$ has very tight bounds ($0.82< \gamma_2 < 0.826280$). We also conjecture that the variance of $|LCS|$ is $cn^2$, with $0.0001<c<0.001$.

As for the distribution of the LCS length. The similar pattern applies for short sequences ($n=2 \sim 15$) applies on the long sequences, with the length of the LCS concentrated in the region $(n-\delta)\pm \delta$. The difference is that $\delta \simeq int(n/\epsilon)+1$, with $\epsilon < 4$, and $\epsilon$ increase linearly with $n$.

\subsection{Multiple sequences}
\label{sec:multi_seq}

There is much theoretical analysis and simulations on the expected length of LCS on two random sequences, in which all characters from $\Sigma$ appear independently. However, (a) there is little research on random sequences with different probabilities for different alphabets, and (b) there is almost no analysis on the expected length of LCS of more than 2 sequences. Here we define ratio $\gamma_k$ for $k$ sequences. Furthermore, we define ratio $\gamma^q_k$ for $k$ sequences on alphabets $q=\Sigma|$.

Theoretically, for $k$ sequences from alphabet set $|\Sigma|, \gamma^q_k=lim_{n\rightarrow\infty}\frac{E(|LCS|)}{n}$ exists because of superadditive properties of LCS.

The lower bounds of $\gamma^q_k$ can be proved by dynamic programming similar to that introduced by Lueker \cite{lueker_improved_2003}. The upper bounds of $\gamma^q_k$ should be at most as big as $\gamma_2$.

Since the exact value of $\gamma^q_k$ is not known, we have analyzed the mean and variances of $\gamma^q_k$ based on Monte-Carlo simulation method.

Based on dynamic programming, the simulations results on the values of $\gamma^2_k$ can be computed for $k=3,4,5$, with the lengths of the sequences $l$ up to 1,000.

We have used $t$ test for evaluation of the mean, and $\chi^2$ test to evaluate the variances. The results are shown in Table \ref{Multi_Seqs}. The average values are tested by $t$ test with $95\%$ confidence, and the variances are tested by $\chi^2$ test with $95\%$ confidence.

\begin{table}[htbp]
\caption{\label{Multi_Seqs} Approximate Length of LCS and its variances for multiple sequences based on $q=2$.}
\begin{center}
\begin{tabular}{|l|l|l|l|l|l|l|}
\hline
k & \multicolumn{2}{|c|}{3} & \multicolumn{2}{|c|}{4} & \multicolumn{2}{|c|}{5} \\
\hline
n & $gamma_k$ & Variance & $gamma_k$ & Variance & $gamma_k$ & Variance \\
\hline
2 & 0.412 & 0.315 & 0.297 & 0.297 & 0.339 & 0.341 \\
\hline
3 & 0.482 & 0.429 & 0.379 & 0.290 & 0.395 & 0.583 \\
\hline
4 & 0.505 & 0.538 & 0.399 & 0.283 & 0.404 & 0.816 \\
\hline
5 & 0.554 & 0.550 & 0.462 & 0.370 & 0.475 & 0.963 \\
\hline
6 & 0.560 & 0.796 & 0.491 & 0.529 & 0.504 & 1.196 \\
\hline
7 & 0.568 & 0.796 & 0.503 & 0.580 & 0.524 & 1.202 \\
\hline
8 & 0.578 & 0.797 & 0.523 & 0.614 & 0.560 & 1.207 \\
\hline
9 & 0.584 & 0.933 & 0.535 & 0.730 & 0.536 & 1.337 \\
\hline
10 & 0.592 & 1.014 & 0.584 & 0.787 & 0.596 & 2.551 \\
\hline
11 & 0.599 & 1.126 & 0.588 & 0.798 & 0.582 & 2.500 \\
\hline
12 & 0.616 & 1.181 & 0.591 & 1.154 & 0.657 & 2.389 \\
\hline
13 & 0.623 & 1.201 & 0.607 & 1.131 & 0.639 & 2.661 \\
\hline
14 & 0.630 & 1.250 & 0.610 & 1.290 & 0.649 & 2.541 \\
\hline
15 & 0.642 & 1.288 & 0.616 & 1.319 & 0.651 & 3.314 \\
\hline
16 & 0.645 & 1.655 & 0.631 & 1.432 & 0.685 & 3.661 \\
\hline
17 & 0.647 & 1.705 & 0.631 & 1.734 & 0.667 & 3.551 \\
\hline
18 & 0.650 & 1.742 & 0.637 & 2.103 & 0.646 & 4.449 \\
\hline
19 & 0.651 & 1.767 & 0.641 & 2.491 & 0.679 & 3.678 \\
\hline
20 & 0.655 & 1.833 & 0.645 & 2.626 & 0.663 & 4.999 \\
\hline
\end{tabular}
\end{center}
\end{table}

The values of $\gamma_k$ increases as $n$ increase, but with slower rate for larger $k$. For $k$ sequences from alphabet set $|\Sigma|$, we conjecture that $\gamma^q_k=\frac{2}{(\frac{q*k}{2})^{0.5}+1}$.

The variances of $|LCS|$, $Var_{|LCS|}$, is also of interest. The variances of $|LCS|$ increase linearly with $n^2$. The variances value also increases as $n$ increase, and with faster rate as $k$ is larger. More specifically, we have observed from simulation results that $Var(|LCS|)=cn^2$, where $c$ is an constant less than 1.

For the distribution of $|LCS|$ for multiple sequences, we have listed some of the results, and the preliminary analysis indicates the approximate natural distribution of $|LCS|$.

\subsection{Non-uniform distribution of alphabets}
\label{sec:non_uniform_alphabet}

More complicated situation is that for different alphabets have different probabilities to appear in the random sequences. The non-random distribution of the alphabets in sequences is normal in real situations, one example is that for coding regions of the DNA sequences, there are more G and C characters than A and T characters ($GC content\geq 50\%$).

There is a conjecture about upper limit in this situation by Mainville that for two sequences on $\Sigma$ alphabets with unequal probabilities $p_1$, $p_2$,...,$p_{\Sigma}$, $\lim_{|n\rightarrow\inf}(\gamma^k)<c_{k*}$, where $k*$ is the greatest integer $\leq 1/\sum(p_i^2)$.

For binary sequences with alphabets {0,1}, let $p$ be the probability that 1 is selected. We define $\gamma_2(p)$ as $\gamma_2$ with probability of alphabet 1 to be $p$. We try to find the asymptotic behavior of $\gamma_2(p)$ as $p$ goes from 0 to 0.5 (due to symmetry, we do not need $p$ from 0.5 to 1). We have done some simulations on this problem.

From results, we have observed the trend that the value of $p$ slowly increases from 0 to 0.5, the value of $\gamma_2(p)$ first increase (from 0.1 to 0.8) rapidly, then it decreases. And the longer the sequences, the more obvious this trend. This comply with the conjecture, but have shows more details on how the values approach the upper limit. The variance of $\gamma_2(p)$ decreases as $p$ increases from 0 to 0.5.

\subsection{Larger $|\Sigma|$}
\label{sec:large_alphabet}

The Sankoff-Mainville conjecture that $\lim_{|\Sigma|\rightarrow\inf}(\gamma_2*\sqrt{|\Sigma|})=2$ is proven to be true. And there are much research on the lower and upper bounds for $E(|LCS|)$ with $|\Sigma|\leq15$. However, there are little research on the analysis of specific larger $|\Sigma|$ (amino acids set and English alphabets set, for example). Moreover, there is no such investigation for multiple sequences.

We have analyzed the mean and variances of $|LCS|$ for larger $|\Sigma|$ based on Monte-Carlo simulation method.

\section{Analysis of heuristic algorithms}
\label{sec:5}

The expected value of $|LCS|$, $\gamma_k$, is not merely of theoretical interest. For heuristic algorithms on LCS problem, $\gamma_k$ can be used as a standard for evaluating their performances. Actually, for two sequences over alphabets set $|\sum|$, Francis Chin et. al. \cite{chin_performance_1994} proved that any heuristic which uses only global information, such as the frequency of symbols, might return a common subsequence as short as $1/|\sum|$ of the length of the longest. Except for a few of such studies, there are few empirical analysis about the evaluation of the heuristic algorithms on LCS problem except for direct results comparison, especially on multiple sequences.

Here we have defined the performance ratio of the heuristic algorithms over an instance $S$ is the value $R_A(S)$, where

\begin{equation}\label{performance_ratio} 
R_A(S)=|opt(S)|/|LCS_A(S)|
\end{equation}

in which $opt(S)$ is the optimal LCS, and $LCS_A(S)$ is the result of algorithm A. The closer performance ratio is to 1, the better the heuristic. For large sequences datasets, $opt(S)$ is generally not available, so that estimation of the upper bound or lower bounds of $opt(S)$ is calculated to assess performance ratios.

In this project, we have compared different algorithms on simulated sequences datasets with different settings, and we have analyzed the means, variances and distributions of results. Results show that for two random sequences, good heuristic algorithms are able to generate common subsequences whose lengths are only $5\%$ to $10\%$ shortere than optimal LCS length, indicating the good performance of these algorithms.

\subsection{Upper Bounds calculation}
\label{sec:UB}

for datasets with many sequences, the dynamic programming approach for the exact LCS is not feasible, so we have computed the upper bound of the LCS length. Given a set $S$ of $n$ sequences with max sequence length $k$, and the alphabet set $\sum = {\sigma_1, \sigma_2, бн, \sigma_{|\sum|}}$.The algorithm first choose number $|\sum|$ of sequences ${S_1,S_2,бнS_{|\sum|}}$ from the sequences set $S$, so that $s_i$ is the sequence with most number of $\sigma_i$. Then dynamic programming is applied on ${S_1,S_2,бнS_{|\sum|}}$, and the upper bound of the LCS length is obtained. If $|\sum|$ is large, or the length of the sequences is very long; then the number of sequences can be smaller than $|\sum|$. These sequences are chosen in the same manner as descried above, but only part of the alphabets in $\sum$ are considered, subject to the alphabet contents of these sequences.

\subsection{Heuristic algorithms for LCS problem}
\label{sec:heuristic_algos}

There exists a {\it dynamic programming algorithm} to compute the LCS of a set $S$ of $k$ sequences, but it requires $O(nk)$ time and space, where $n$ is the length of the longest sequence in S. Hence this algorithm is feasible only for small values of $n$ and $k$.

The simple and fast {\it Long Run algorithm} is proposed by Jiang and Li \cite{jiang_approximation_1994}. For $k$ sequences in $S = {s_1, s_2, бн, s_k}$ on the finite alphabet set $\sum = {\sigma_1, \sigma_2, бн, \sigma_{|\sum|}}$. Long Run algorithm finds maximum $m$ such that there is $\sigma_i$ in $\sum$, and $\sigma_im$ is a common subsequence of all input sequences. It outputs $\sigma_im$ as the result of LCS. The time complexity of the long run algorithm is $O(kn)$.

These are also two variations of an iterative scheme based on combining "best" sequence pairs. Given any pair of sequences, $S_i$ and $S_j$, an optimal subsequence of the pair, $LCS(S_i,S_j)$, can be computed in $O(n^2)$ using dynamic programming. The {\it Greedy algorithm} first chooses the "best" sequence pair - the pair that gives the longest $LCS(S_i,S_j)$. Let's call them $S_1$ and $S_2$. The algorithm then replaces the two sequences $S_1$ and $S_2$ by their subsequence, $LCS(S_1,S_2)$. The algorithm proceeds recursively. Thus, we can express it as follows: 

\begin{equation}\label{Greedy_algo}
Greedy(S_1,S_2,бн, S_k) = Greedy(LCS(S_1,S_2),S_3,бн,S_k)
\end{equation}

The {\it Tournament algorithm} is similar to the Greedy algorithm. It builds a "tournament" based on finding multiple best pairs at each round and can be expressed schematically as follows: 

\begin{equation}\label{Tournament_algo}
Tournament(S_1,S_2,бн,S_k) = Tournament((LCS(S_1,S_2),LCS(S_3,S_4),бн, LCS(S_{k-1},S_k)))
\end{equation}

Both Greedy and Tournament algorithms have $O(k^2n^2)$ time complexity and $O(kn + n^2$) space complexity.

To tackle the LCS problem, we have proposed the {\it Deposition and Extension algorithm} \cite{ning_deposition_2009}, in which a common subsequence for a set of sequences is first generated based on searching for common characters in a certain range of every sequences, then these common characters are concatenated to form a common subsequence, and subsequenctly extended to get the result. This algorithm is based on character-by-character approach.

Based on the results on simulated sequences, we have found that for two sequences, the heuristic algorithms can generate common subsequences whose lengths are about $5\%$ to $10\%$ shorter than the optimal LCS lengths, which are currently best heuristic results. The Deposition and Extension algorithm is currently one of the best heuristic algorithms in terms of efficiency for the LCS problem.

\subsection{Performance analysis}
\label{sec:performance_analysis}

We have compared these algorithms by their performance ratios based on simulated and real sequences \cite{ning_deposition_2009}. Results show that Deposition and Extension algorithm has the best performance ratio on most of datasets, followed by Long Run algorithm. The Tournament algorithm and Greedy algorithm performs worst, especially when there are many sequences in the datasets.

\section{Discussions}
\label{sec:5}

In this paper, we have assessed the expected values, variants and distribution of LCS for two or more sequences over different settings systematically. Specificcly, seuqences set (a) with multiple seuqences, (b) based on multiple alphabets and (c) with uneven alpahbet probabilities are analyzed. Results show that though there's trands that can be observed about expected values, variants and distribution, these values different greatly for different sequences sets.

We have also analyze the performance ratios of current heurisitc algorithms based on these systematic analysis results as benchmark. Results have shown that Deposition and Extension algorithm has the best performance.

\bibliographystyle{elsart-num}
\bibliography{Simplify(LCS)}

\end{document}